\documentclass[runningheads]{llncs}

\ifdefined\pdfpagewidth
\fi
\usepackage[T1]{fontenc}
\usepackage{amsmath,amssymb,amsfonts}
\usepackage{graphicx}
\usepackage[table]{xcolor}
\usepackage{booktabs}
\usepackage{colortbl}
\usepackage{adjustbox}
\usepackage{url}
\usepackage{microtype}
\usepackage{bbding}
\usepackage{hyperref}
\hypersetup{hidelinks}

\begin{document}

\title{IMEX-FND: A Traceable Interaction-Aware Mixture-of-Experts Framework for Multimodal Fake News Detection}
\titlerunning{IMEX-FND}

\author{Yuchen Miao\inst{1} \and Zijun Wang\inst{1}\textsuperscript{\Envelope} \and Ke Liu\inst{1} \and Peixuan Wang\inst{2} \and \mbox{Chang Han\inst{1}}}
\authorrunning{Y. Miao et al.}
\institute{Sydney Smart Technology College, Northeastern University, China\\
\and
School of Computer and Communication Engineering, Northeastern University at Qinhuangdao, China\\
\textsuperscript{\Envelope}Corresponding author.}

\maketitle
\thispagestyle{headings}

\begin{abstract}
Multimodal fake news detection (FND) increasingly demands verdicts that are not only accurate but traceable---revealing how cross-modal evidence is combined---yet two coupled difficulties remain. First, text--image relations are heterogeneous: uniqueness, redundancy, and synergy coexist and vary from post to post, so a single global fusion rule is brittle and opaque. Second, the dominant modality shifts across instances, which static encoders and a fixed fusion pathway handle poorly. We present IMEX-FND, an interaction-aware mixture-of-experts framework that couples adaptive routing with explicit interaction decomposition. A Multi-Modal Expert Gateway (MMEG) performs instance-wise, within-modality routing over specialized and shared experts and builds a CLIP-grounded cross-modal stream, yielding three refined, interaction-ready representations that adapt to the dominant modality of each post. An Interaction-aware Multi-Modal Expert Fusion (IMEF) module then decomposes the interactions among these streams into \emph{uniqueness}, \emph{redundancy}, and \emph{synergy}, producing transparent, sample-wise weights via a modality-replacement training signal. The two stages form a single route-then-decompose pipeline whose routing and interaction weights are both inspectable, offering instance- and dataset-level traceability for misinformation diagnosis. Extensive experiments on the Weibo, Weibo-21, and Gossip benchmarks show that IMEX-FND achieves state-of-the-art performance, surpassing competitive baselines by 0.4--1.2\% while offering superior traceability with fewer parameters.
\end{abstract}

\section{Introduction}
\label{sec:intro}

Online social platforms have become a primary channel for producing and consuming news; while democratizing information, they also lower the cost of fabricating and amplifying deceptive content at scale. Consequently, fake news has become a persistent societal risk, as intentionally false or misleading content can be crafted to manipulate beliefs and behavior, motivating a long line of automated detection methods.

\begin{figure}[t]
  \centering
  \includegraphics[width=0.7\linewidth, page=1, trim=0 0 0 0, clip]{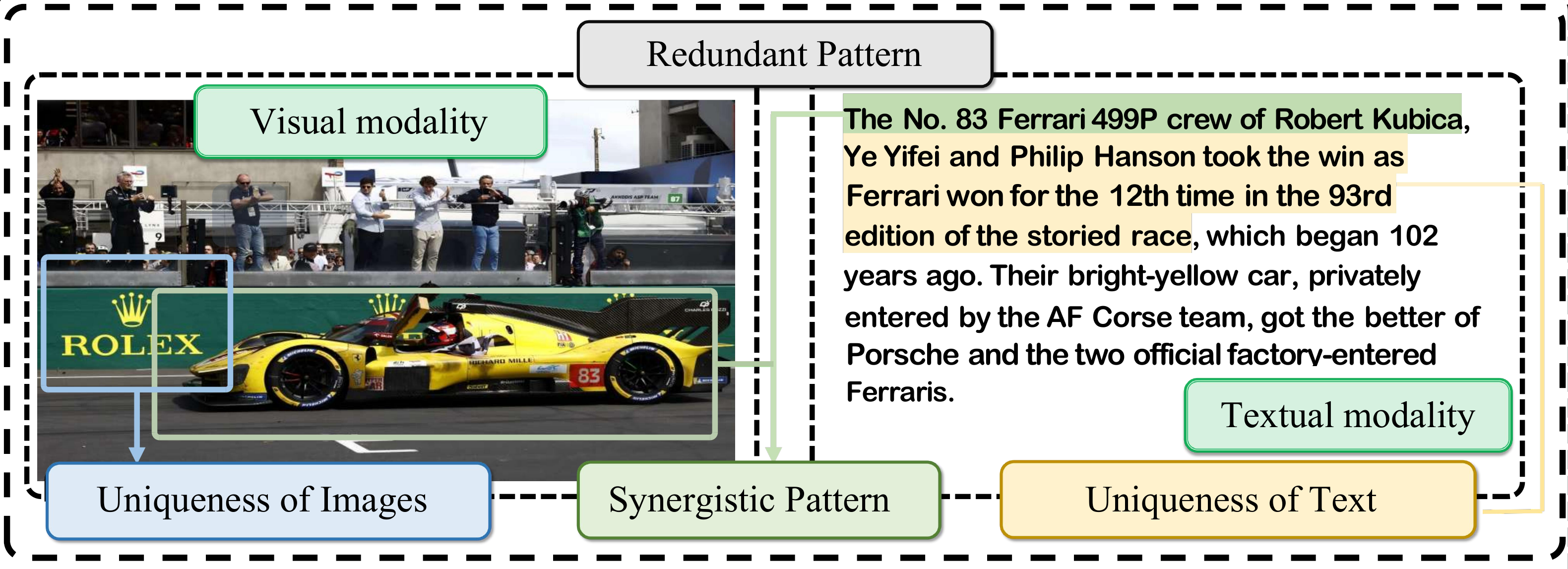}
  \caption{An interaction-aware view of multimodal fusion showing redundant (duplicate) and modality-unique information.}
  \label{fig:intro-limitations}
  \vspace{-1em}
\end{figure}

Most deceptive posts are multimodal: a textual claim is paired with an image that may corroborate, complement, or contradict it. As Figure~\ref{fig:intro-limitations} shows, this text--image relation is heterogeneous: some evidence is \emph{unique} to one modality, some is \emph{redundantly shared}, and some emerges only from their combination (\emph{synergy}). Two difficulties follow. First, the interaction that exposes a fake differs from post to post, so a single global fusion rule that compresses text and image into one vector is brittle and opaque about why a verdict is reached. Second, the dominant modality shifts across instances -- one post is betrayed by its text, another by its image -- and a static encoder with a fixed fusion pathway cannot adapt, capping robustness.

Existing methods tackle parts of this picture but not the whole. Late-fusion pipelines concatenate or co-attend over text--image features atop pretrained encoders \cite{jin2017multimodal,khattar2019mvae}; consistency- and ambiguity-driven methods regularize cross-modal coherence via similarity, ambiguity, or contrastive objectives \cite{zhou2020safe,C3,wang2023crossmodalcontrastive,zhou2023clipguided}; and expert-based designs use gating, domain adaptation, or mixture-of-experts (MoE) routing to pick a fusion pathway per instance \cite{yu2023mmoe,tong2024mmdfnd,lu2025dammfnd,tong2025dapt,lu2026blind,C9}. Yet ambiguity-guided methods still collapse interaction complexity into a scalar weight, and learned experts and routes are rarely interpretable as canonical interaction types -- leaving both robustness under shifting modalities and decision transparency unresolved.

We propose IMEX-FND, an interaction-aware mixture-of-experts framework that makes modality interactions first-class through a two-stage route-then-decompose design. A Multi-Modal Expert Gateway (MMEG) routes each modality instance-wise through specialized and shared experts and grounds a cross-modal stream in vision--language pretraining \cite{zhou2023clipguided}, yielding three refined representations (text, image, fusion) that adapt to whichever modality carries the decisive evidence. An Interaction-aware Multi-Modal Expert Fusion (IMEF) then decomposes their interactions into human-readable \emph{uniqueness}, \emph{redundancy}, and \emph{synergy}, emitting transparent, sample-wise weights learned with a modality-replacement signal that drives each expert to specialize \cite{liang2023factorized,Xin2025I2MoE}.

The main contributions of this paper are summarized as follows:
\begin{itemize}
\item We propose IMEX-FND, a unified interaction-aware MoE framework coupling instance-adaptive modality routing with explicit interaction decomposition for multimodal FND.
\item We design a Multi-Modal Expert Gateway (MMEG) that refines each modality through specialized and shared experts and builds a CLIP-grounded fusion stream, improving robustness under dominant-modality shift.
\item We design a traceable Interaction-aware Multi-Modal Expert Fusion (IMEF) that disentangles uniqueness, redundancy, and synergy into sample-wise weights via modality-replacement weak supervision, yielding instance- and dataset-level interpretability.
\item On Weibo, Weibo-21, and Gossip, IMEX-FND attains state-of-the-art accuracy with fewer parameters while exposing interpretable interaction attributions.
\end{itemize}

\section{Related Work}
\label{sec:related_work}

\subsection{Modality Interactions}
Two lines of work underpin our design: how modalities interact, and how multimodal fake news detectors exploit that interaction. Modality-interaction research studies how modalities jointly support inference, where effective reasoning often leverages both unimodal and cross-modal evidence. Factorized contrastive learning captures interaction factors from an information-theoretic view \cite{liang2023factorized}, while gated routing models diverse interaction patterns with separate learning pathways \cite{yu2023mmoe,Xin2025I2MoE}. In multimodal fake news detection, models should exploit both unique and shared evidence, yet many do not explicitly model semantic-level interaction types, yielding brittle fusion under heterogeneous relations. We address this by explicitly modeling interaction patterns and exposing transparent, instance-wise fusion behavior.

\subsection{Multimodal Fake News Detection}
Early work combines modalities with attention and sequence models \cite{jin2017multimodal}, while MVAE learns shared representations via variational objectives \cite{khattar2019mvae}. With pretrained encoders, later methods concatenate or co-attend over text--image features, though cross-modal misalignment can still harm performance.

To improve coherence, many approaches add consistency, ambiguity modeling, or contrastive objectives \cite{C3,wang2023crossmodalcontrastive,zhou2020safe}. SAFE estimates cross-modal inconsistency by similarity \cite{zhou2020safe}, while CAFE quantifies ambiguity via KL-based divergence between modality distributions \cite{C3}, and CMC distills cross-modal correlations implicitly \cite{wei2022crossmodalkd}. COOLANT integrates cross-modal contrastive learning with ambiguity-aware fusion \cite{wang2023crossmodalcontrastive}, and CLIP-guided fusion exploits vision--language pretraining for stronger semantic alignment signals \cite{zhou2023clipguided}. Recent extensions inject high-level signals, such as intent--semantic joint learning via graph modeling \cite{wang2025bridgingthoughtswordsgraphbased}, external knowledge augmentation with emotion guidance \cite{zhu2025kenknowledgeaugmentationemotion}, and LLM-augmented reinforced sampling to strengthen supervision and hard-example coverage \cite{tong-etal-2025-generate}. Beyond ambiguity-weighted aggregation, MIMoE-FND routes instances to specialized fusion experts by supervising a hierarchical MoE gate with unimodal prediction agreement and semantic alignment \cite{C9}. More recently, multimodal large language models (MLLMs) enable explicit reasoning: DIVER performs iterative visual evidence reasoning, invoking fine-grained visual tools only when text is insufficient \cite{zhou2026diver}, while multi-perspective rationale generation with cross-verification reconciles contradictions among MLLM-generated rationales before prediction \cite{chen2026rationale}. In parallel, MViR strengthens the visual branch with multi-view visual--semantic representations via pyramid dilated convolution \cite{liang2026mvir}. Retrieval-augmented detection and multimodal fact-checking resources further emphasize external evidence and retrieval difficulty \cite{li2026retrieval,xu2025mmm}.

Despite progress, ambiguity-guided methods compress interaction complexity into scalar weights, with limited insight into which interaction dominates and why, while MoE routing improves adaptivity yet rarely interprets experts as canonical interaction types. We therefore introduce an interaction-aware, traceable fusion design that disentangles heterogeneous interactions and provides transparent, instance-level attributions for robust detection.

\section{Methodology}
\label{sec:method}

\begin{figure}[t]
  \centering
  \includegraphics[page=1,width=\textwidth]{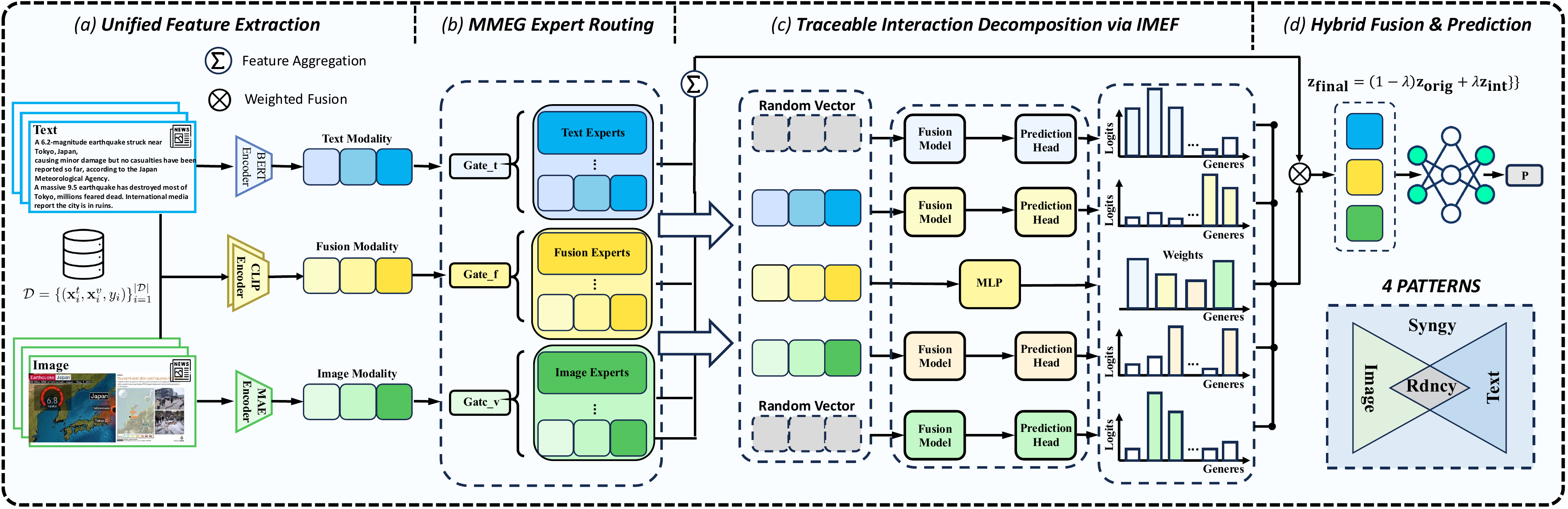}
  \caption{Overall framework. Pretrained encoders extract token-level representations. MMEG routes each modality through a mixture of specialized and shared experts to obtain modality features. IMEF explicitly models interaction types via traceable expert weights and weakly-supervised interaction losses. Final prediction uses hybrid fusion with multi-auxiliary supervision.}
  \label{fig:fnd}
\end{figure}

\subsection{Problem Setup}
We study multimodal fake news detection with a dataset
$\mathcal{D}=\{(\mathbf{x}_i^{t},\mathbf{x}_i^{v}, y_i)\}_{i=1}^{|\mathcal{D}|}$,
where $\mathbf{x}^{t}$ is a text sequence, $\mathbf{x}^{v}$ is an associated image (or a representative image for multi-image posts), and $y\in\{0,1\}$ indicates fake vs.\ real. Our goal is to learn a predictor $f(\mathbf{x}^{t},\mathbf{x}^{v})\rightarrow p(y{=}1)$.

\subsection{Overview}
IMEX-FND is an interaction-aware mixture-of-experts framework with four parts that turn a raw post into a traceable verdict. Pretrained encoders first provide text, image, and cross-modal signals. A Multi-Modal Expert Gateway (MMEG) then mixes experts within each modality into compact, modality-specific features, and an Interaction-aware Multi-Modal Expert Fusion (IMEF) module decomposes their interaction into \emph{uniqueness}, \emph{synergy}, and \emph{redundancy} experts that emit sample-wise traceable weights. Multi-auxiliary supervision with a hybrid fusion objective then ties the pieces together, forming a single route-then-decompose pipeline whose intermediate weights stay inspectable. The pretrained backbones serve as frozen feature extractors, while the MoE modules and classifiers are trained end-to-end.

\subsection{Pretrained Encoders and Unified Representations}
Each post is first turned into three signals that feed MMEG. The text encoder produces token features $\mathbf{T}_{\mathrm{raw}}\in\mathbb{R}^{B\times L\times d_t}$ together with a routing summary obtained by masked attention,
\begin{equation}
\mathbf{t}_{\mathrm{pool}} = \mathrm{Attn}_{\mathrm{mask}}(\mathbf{T}_{\mathrm{raw}})\in\mathbb{R}^{B\times d_t},
\end{equation}
and, symmetrically, the image encoder produces patch features $\mathbf{I}_{\mathrm{raw}}\in\mathbb{R}^{B\times N\times d_v}$ with a token-attention summary,
\begin{equation}
\mathbf{v}_{\mathrm{pool}} = \mathrm{Attn}(\mathbf{I}_{\mathrm{raw}})\in\mathbb{R}^{B\times d_v}.
\end{equation}
To capture cross-modal evidence directly, we also extract CLIP image and text features $\mathbf{c}^{v},\mathbf{c}^{t}\in\mathbb{R}^{B\times d_c}$ and project their concatenation,
\begin{equation}
\mathbf{c} = \phi\big([\mathbf{c}^{v};\mathbf{c}^{t}]\big)\in\mathbb{R}^{B\times d},
\end{equation}
where $\phi(\cdot)$ is an MLP projection and $d$ is the unified feature dimension shared by all downstream experts (see Sec.~\ref{subsubsec:impl_details}).

\subsection{Multi-Modal Expert Gateway (MMEG)}
\label{subsec:mmeg_revised}
\begin{figure}[t]
  \centering
  \includegraphics[width=0.7\linewidth]{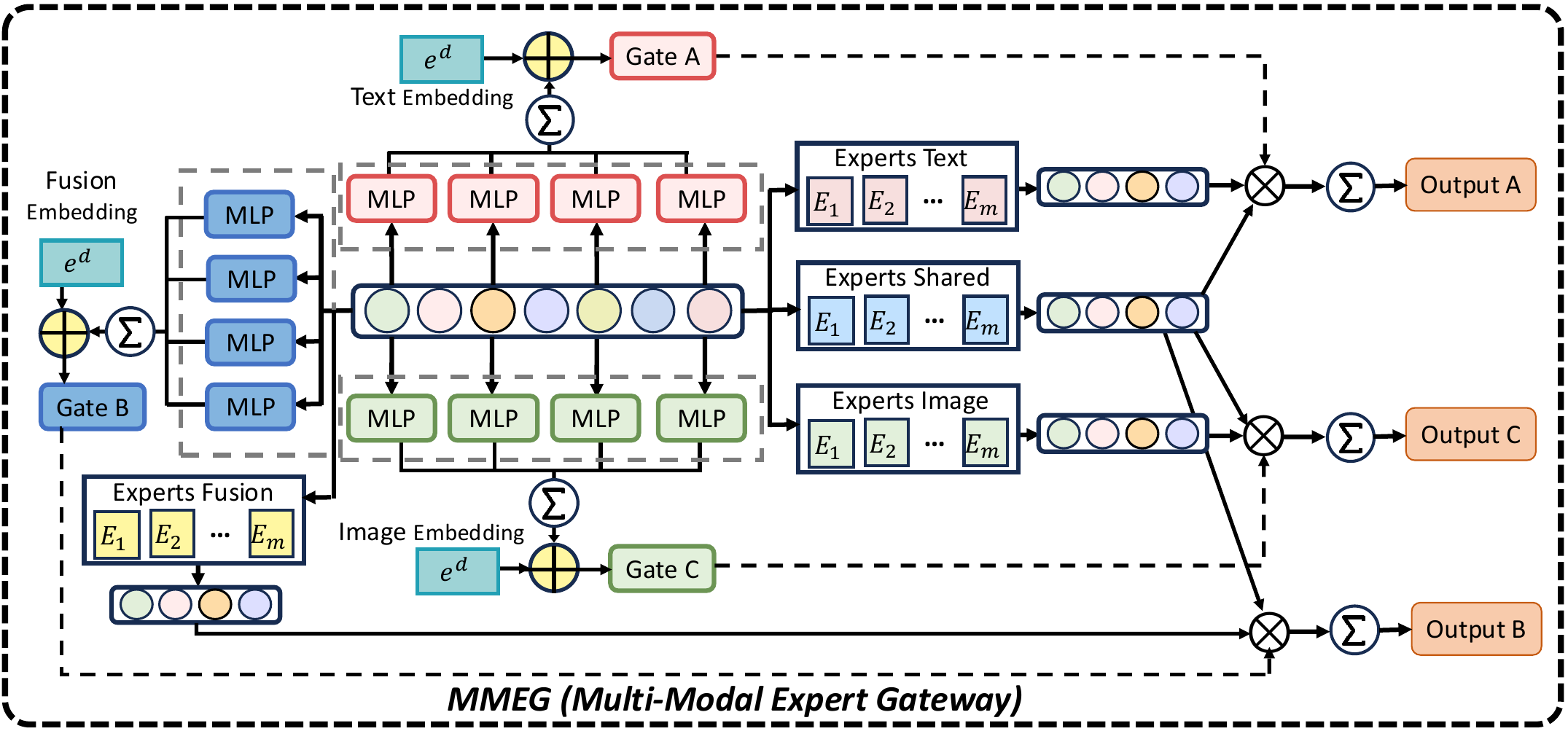}
  \caption{Detailed architecture of the Multi-Modal Expert Gateway (MMEG). Each modality is routed by its own gate (Gate~A/B/C for the text/fusion/image streams) over a pool of specialized experts together with a set of shared experts; the gated, summed combination of expert outputs yields the three refined streams (Output~A/B/C, i.e.\ $\mathbf{t}$, $\mathbf{m}$, $\mathbf{v}$). The fusion stream is constructed from the CLIP-grounded embedding and the shared evidence of the text and image streams.}
  \label{fig:mmeg}
\end{figure}
Following the expert-routing paradigm of multi-gate MoE and layered expert extraction \cite{yu2023mmoe,tong2024mmdfnd}, MMEG performs within-modality routing (Figure~\ref{fig:mmeg}) that refines each representation while preserving its modality-specific bias. For text and image, each specialized expert is a multi-kernel 1D-CNN extractor over the token sequence,
\begin{equation}
E(\mathbf{X}) = \mathrm{Concat}\Big(\max\nolimits_{l}\mathrm{Conv}_{k}(\mathbf{X})\Big)_{k\in\{1,2,3,5,10\}}
\in\mathbb{R}^{B\times d},
\end{equation}
with max pooling over the sequence length; shared experts tie parameters across text and image, and fusion experts are lightweight MLP refiners in $\mathbb{R}^{d}$. Given a pooled routing input $\mathbf{u}^m$ for each modality $m\in\{T,I,F\}$, a gate produces instance-wise expert weights,
\begin{equation}
\mathbf{w}^m = \mathrm{Softmax}\!\Big(\mathrm{MLP}_{\mathrm{gate}}^m(\mathbf{u}^m)\Big)\in\mathbb{R}^{K+L},
\end{equation}
where $K$ and $L$ count the specialized and shared experts (see Sec.~\ref{subsubsec:impl_details}); text and image take $\mathbf{u}^T{=}\mathbf{t}_{\mathrm{pool}}$ and $\mathbf{u}^I{=}\mathbf{v}_{\mathrm{pool}}$ and operate on the token sequences $\mathbf{X}^T{=}\mathbf{T}_{\mathrm{raw}}$ and $\mathbf{X}^I{=}\mathbf{I}_{\mathrm{raw}}$. Each modality's routed feature is then the gated combination of its experts,
\begin{equation}
\mathbf{H}^m=\sum_{k=1}^{K}w_k^m\,E_k^m(\mathbf{X}^m)+\sum_{l=1}^{L}w_{K+l}^m\,S_l(\mathbf{X}^m)\in\mathbb{R}^{B\times d}.
\end{equation}
The fusion modality is built by combining the CLIP cross-modal feature with the shared evidence of the two unimodal streams,
\begin{equation}
\mathbf{z}_{\mathrm{fus}} = \psi\big([\mathbf{c}; \mathbf{H}_{\mathrm{sh}}^T; \mathbf{H}_{\mathrm{sh}}^I]\big)\in\mathbb{R}^{B\times d},
\end{equation}
where $\mathbf{H}_{\mathrm{sh}}^T,\mathbf{H}_{\mathrm{sh}}^I$ are shared-expert outputs and $\psi(\cdot)$ is an MLP. A fusion gate finally yields $\mathbf{H}^F$, giving the three MMEG outputs consumed by the next stage,
\begin{equation}
\mathbf{t}=\mathbf{H}^T,\qquad \mathbf{v}=\mathbf{H}^I,\qquad \mathbf{m}=\mathbf{H}^F.
\end{equation}

\subsection{Interaction-aware Multi-Modal Expert Fusion (IMEF)}
\label{subsec:imef_revised}
Given the three MMEG streams, IMEF decomposes their interaction into a taxonomy inspired by partial information decomposition \cite{liang2023factorized,Xin2025I2MoE}: \emph{uniqueness} (text, image, fusion), \emph{synergy}, and \emph{redundancy}. Stacking the streams into $\mathbf{z}=[\mathbf{t};\mathbf{v};\mathbf{m}]\in\mathbb{R}^{B\times 3d}$, five interaction experts $\{\mathcal{E}_{T},\mathcal{E}_{I},\mathcal{E}_{F},\mathcal{E}_{\mathrm{syn}},\mathcal{E}_{\mathrm{red}}\}$, each a 2-layer MLP, read this joint vector,
\begin{equation}
\mathcal{E}_i(\mathbf{z})=\mathrm{MLP}_i(\mathbf{z})\in\mathbb{R}^{B\times d},
\end{equation}
and we write $\mathbf{o}_i^{(0)}=\mathcal{E}_i(\mathbf{z})$ for the output on the unperturbed input. A reweighting network maps this vector to per-sample interaction weights,
\begin{equation}
\boldsymbol{\alpha}=\mathrm{Softmax}\!\Big(\mathrm{MLP}_{\mathrm{rw}}(\mathbf{z})/\tau\Big)\in\mathbb{R}^{B\times 5},
\end{equation}
with temperature $\tau$ controlling sharpness, and the fused interaction feature is their weighted sum,
\begin{equation}
\mathbf{z}_{\mathrm{int}}=\sum_{i=1}^{5}\alpha_i\,\mathbf{o}_i^{(0)}\in\mathbb{R}^{B\times d}.
\end{equation}
Because $\boldsymbol{\alpha}$ is emitted per sample, it doubles as an attribution that can be aggregated for global interaction profiling.

To make each expert specialize in its interaction type, we add a weakly-supervised modality-replacement signal, applied only during training. For $r\in\{T,I,F\}$ we overwrite one stream with a noise vector $\mathbf{n}_r$ while keeping the others intact,
\begin{equation}
\tilde{\mathbf{z}}^{(r)}=
\big[\tilde{\mathbf{t}};\tilde{\mathbf{v}};\tilde{\mathbf{m}}\big],\quad
\tilde{\mathbf{h}}=
\begin{cases}
\mathbf{n}_r, & \text{if }\mathbf{h}\text{ corresponds to modality }r,\\
\mathbf{h}, & \text{otherwise},
\end{cases}
\end{equation}
with $(\mathbf{h},\tilde{\mathbf{h}})\in\{(\mathbf{t},\tilde{\mathbf{t}}),(\mathbf{v},\tilde{\mathbf{v}}),(\mathbf{m},\tilde{\mathbf{m}})\}$, and pass each perturbed input through the experts to obtain $\mathbf{o}_i^{(r)}=\mathcal{E}_i(\tilde{\mathbf{z}}^{(r)})$. Each interaction type then receives a loss matching its meaning. The \emph{uniqueness} expert of modality $r$ should react only when $r$ is removed, so we use a triplet margin with the unperturbed output as anchor, the $r$-replaced output as negative, and the remaining replacements as positives,
\begin{equation}
\mathcal{L}_{\mathrm{uniq}}^{(r)}
=\frac{1}{2}\sum_{r'\in\{T,I,F\}\setminus\{r\}}
\ell_{\mathrm{tri}}\Big(\mathbf{o}_{r}^{(0)},\ \mathbf{o}_{r}^{(r')},\ \mathbf{o}_{r}^{(r)}\Big),
\end{equation}
where $\ell_{\mathrm{tri}}(\mathbf{a},\mathbf{p},\mathbf{n})=\max\big(0,\|\mathbf{a}-\mathbf{p}\|_2-\|\mathbf{a}-\mathbf{n}\|_2+\mu\big)$ and $\mu$ is the margin. The \emph{synergy} expert, in contrast, should collapse whenever any modality is missing, so we penalize its similarity to each perturbed output,
\begin{equation}
\mathcal{L}_{\mathrm{syn}}
=\frac{1}{3}\sum_{r\in\{T,I,F\}}
\mathrm{cos}\big(\mathbf{o}_{\mathrm{syn}}^{(0)},\mathbf{o}_{\mathrm{syn}}^{(r)}\big),
\end{equation}
whereas the \emph{redundancy} expert should stay invariant to single-modality replacement, which we encourage with the complementary objective
\begin{equation}
\mathcal{L}_{\mathrm{red}}
=\frac{1}{3}\sum_{r\in\{T,I,F\}}
\Big(1-\mathrm{cos}\big(\mathbf{o}_{\mathrm{red}}^{(0)},\mathbf{o}_{\mathrm{red}}^{(r)}\big)\Big).
\end{equation}
Summing the five terms gives the interaction regularizer
\begin{equation}
\mathcal{L}_{\mathrm{int}}
=\mathcal{L}_{\mathrm{uniq}}^{(T)}+\mathcal{L}_{\mathrm{uniq}}^{(I)}+\mathcal{L}_{\mathrm{uniq}}^{(F)}
+\mathcal{L}_{\mathrm{syn}}+\mathcal{L}_{\mathrm{red}}.
\end{equation}
Replacement is confined to training; at inference IMEF uses only $\mathbf{z}_{\mathrm{int}}$ from the unperturbed input.

\subsection{Prediction Heads and Hybrid Fusion Objective}
\label{subsec:objective_revised}
Three lightweight MLP classifiers attach auxiliary predictions to the per-modality streams,
\begin{equation}
p_T=\sigma(C_T(\mathbf{t})),\quad
p_I=\sigma(C_I(\mathbf{v})),\quad
p_F=\sigma(C_F(\mathbf{m})),
\end{equation}
while the final decision blends the summed representation $\mathbf{z}_{\mathrm{orig}}=\mathbf{t}+\mathbf{v}+\mathbf{m}$ with the interaction feature,
\begin{equation}
\mathbf{z}_{\mathrm{final}}=(1-\lambda)\mathbf{z}_{\mathrm{orig}}+\lambda \mathbf{z}_{\mathrm{int}},
\end{equation}
where $\lambda\in[0,1]$ controls the contribution of interaction-aware fusion (see Sec.~\ref{subsubsec:impl_details}). The blended feature is classified by a final head,
\begin{equation}
p_{\mathrm{fake}}=\sigma(C(\mathbf{z}_{\mathrm{final}})).
\end{equation}

The full model is trained end-to-end by minimizing the total objective
\begin{equation}
\begin{aligned}
\mathcal{L}_{\mathrm{total}}
&=\underbrace{\mathcal{L}_{\mathrm{main}}}_{\mathrm{BCE}(p_{\mathrm{fake}},y)}
+\frac{1}{3}\Big(
\underbrace{\mathcal{L}_{T}}_{\mathrm{BCE}(p_T,y)}
+\underbrace{\mathcal{L}_{I}}_{\mathrm{BCE}(p_I,y)}
+\underbrace{\mathcal{L}_{F}}_{\mathrm{BCE}(p_F,y)}
\Big)\\
&\quad
+\sum_{b=1}^{5} w_b\,\mathcal{L}_{\mathrm{int}}^{(b)}
+\delta\lVert\Theta\rVert_2^2 .
\end{aligned}
\end{equation}
where $\{\mathcal{L}_{\mathrm{int}}^{(b)}\}_{b=1}^{5}$ correspond to $\mathcal{L}_{\mathrm{uniq}}^{(T)}$, $\mathcal{L}_{\mathrm{uniq}}^{(I)}$, $\mathcal{L}_{\mathrm{uniq}}^{(F)}$, $\mathcal{L}_{\mathrm{syn}}$, and $\mathcal{L}_{\mathrm{red}}$.
We use fixed interaction weights $\{w_b\}_{b=1}^{5}$ and weight decay $\delta$; hyperparameters are given in Sec.~\ref{subsubsec:impl_details}.

\section{Experiments and Results}
\label{sec:typestyle}

\subsection{Experimental Settings}

\subsubsection{Implementation Details}
\label{subsubsec:impl_details}
\noindent We freeze all pretrained backbones: a BERT-style text encoder, an MAE-pretrained ViT-B/16 for images, and CLIP ViT-B/16 for cross-modal features. Text is truncated to at most $197$ tokens and images are resized to $224{\times}224$ with patch size $16$ (196 patches plus a [CLS] token), and every stream is projected to a unified dimension $d{=}320$. Within MMEG, each modality uses $K{=}6$ specialized and $L{=}12$ shared experts, the multi-kernel CNN has kernel sizes $\{1,2,3,5,10\}$ with 64 filters each, and the routing gate is a 2-layer SiLU MLP with dropout $0.1$; IMEF's five interaction experts are 2-layer MLPs (hidden $320$, dropout $0.1$) and its reweighting network is a 2-layer MLP (hidden $256$, $\tau{=}1.0$), while all classifier heads are 2-layer MLPs (hidden $384$, dropout $0.2$). We optimize with Adam (learning rate $1{\times}10^{-4}$, weight decay $5{\times}10^{-5}$) at batch size $64$ for up to $50$ epochs with early stopping, using BCE for the main and auxiliary heads (auxiliary losses averaged), a triplet margin $\mu{=}1.0$ for uniqueness and cosine penalties for synergy and redundancy, with all interaction losses sharing weight $w_b{=}0.01$ and a hybrid-fusion weight $\lambda{=}0.3$.

\subsubsection{Baselines}
\label{subsec:baselines}
We compare our method against baselines from three paradigm families:
(i) Representation and Robustness Learning, including the domain-robust model DAMMFND~\cite{lu2025dammfnd}, cross-modal distillation frameworks CMC~\cite{wei2022crossmodalkd} and BMR~\cite{ying2023bootstrapping}, and the recent multi-view visual-representation method MViR~\cite{liang2026mvir};
(ii) Cross-Modal Consistency and Alignment, covering SAFE~\cite{zhou2020safe}, CAFE~\cite{C3}, foundation-model-based FND-CLIP~\cite{zhou2023clipguided}, and expert-based MIMoE-FND~\cite{C9};
and (iii) Advanced Semantics and Reasoning, representing the latest state-of-the-art methods: GSFND~\cite{tong-etal-2025-generate}, KEN~\cite{zhu2025kenknowledgeaugmentationemotion}, and InSide~\cite{wang2025bridgingthoughtswordsgraphbased}.

\subsection{Overall Performance Comparison}
\label{ssec:subhead}

\definecolor{bestcolor}{RGB}{255,217,178} 
\definecolor{secondcolor}{RGB}{218,232,252} 
\newcommand{\best}[1]{\cellcolor{bestcolor}\textbf{#1}}
\newcommand{\second}[1]{\cellcolor{secondcolor}\textbf{#1}}

\begin{table}[htbp]
  \centering
  \caption{Performance on \textit{Weibo}, \textit{Weibo-21}, and \textit{Gossip}. Results are averaged over 10 runs; shaded cells indicate the best (\colorbox{bestcolor}{red}) and second-best (\colorbox{secondcolor}{blue}) values in each column. For compactness, leading zeros are omitted for values below one.}
  \label{tab:overall}
  \begingroup
  \renewcommand{\arraystretch}{1.25}
  \setlength{\tabcolsep}{1.0pt}
  \scriptsize
  \begin{adjustbox}{width=\textwidth}
  \begin{tabular}{l *{21}{c}}
  \toprule
  & \multicolumn{7}{c}{\textit{Weibo}} & \multicolumn{7}{c}{\textit{Weibo-21}} & \multicolumn{7}{c}{\textit{Gossip}} \\
  \cmidrule(lr){2-8}\cmidrule(lr){9-15}\cmidrule(lr){16-22}
  Method & Acc & \multicolumn{3}{c}{Fake} & \multicolumn{3}{c}{Real} & Acc & \multicolumn{3}{c}{Fake} & \multicolumn{3}{c}{Real} & Acc & \multicolumn{3}{c}{Fake} & \multicolumn{3}{c}{Real} \\
  \cmidrule(lr){3-5}\cmidrule(lr){6-8}\cmidrule(lr){10-12}\cmidrule(lr){13-15}\cmidrule(lr){17-19}\cmidrule(lr){20-22}
  & & P & R & F1 & P & R & F1 & & P & R & F1 & P & R & F1 & & P & R & F1 & P & R & F1 \\
  \midrule
  EANN & .825 & .845 & .810 & .827 & .805 & .841 & .823 & .868 & .900 & .823 & .860 & .839 & .910 & .873 & .862 & .700 & .515 & .593 & .885 & .954 & .918 \\
  SAFE & .760 & .829 & .722 & .772 & .693 & .809 & .747 & .903 & .891 & .906 & .898 & .914 & .899 & .906 & .836 & .756 & .556 & .641 & .855 & .935 & .893 \\
  SpotFake & .890 & .900 & \best{.964} & .931 & .845 & .654 & .737 & .850 & .951 & .731 & .827 & .784 & \best{.964} & .865 & .856 & .730 & .370 & .491 & .864 & .961 & .910 \\
  CAFE & .838 & .853 & .828 & .840 & .823 & .849 & .836 & .880 & .855 & .913 & .883 & .905 & .842 & .872 & .865 & .730 & .488 & .585 & .885 & .955 & .919 \\
  CMC & .871 & .880 & .865 & .872 & .843 & .858 & .850 & .893 & .903 & .888 & .895 & .881 & .896 & .888 & .881 & .750 & .558 & .640 & .898 & \second{.961} & .928 \\
  BMR & .916 & .880 & \second{.948} & .913 & \best{.942} & .877 & .908 & .927 & .906 & .947 & .926 & .944 & .904 & .924 & .893 & .750 & .637 & .689 & .919 & \best{.965} & .934 \\
  FND-CLIP & .905 & .912 & .899 & .905 & .912 & .899 & .905 & \second{.943} & .935 & .945 & .940 & \best{.950} & .942 & \second{.946} & .878 & .759 & .547 & .636 & .897 & .957 & .926 \\
  DAMMFND & .907 & .921 & .906 & .913 & .891 & .906 & .898 & .931 & .948 & .933 & .940 & .919 & .934 & .926 & .891 & .756 & .648 & .700 & .914 & .960 & .937 \\
  MIMoE-FND & .926 & \second{.938} & .923 & .930 & .913 & .928 & .920 & .941 & \second{.956} & .941 & \second{.948} & .929 & .944 & .936 & .896 & .762 & .644 & .698 & .918 & .962 & .939 \\
  InSide & .881 & .720 & .651 & .684 & .925 & .939 & .932 & .910 & .900 & .910 & .905 & .912 & .918 & .915 & \second{.900} & \second{.785} & \second{.809} & \second{.797} & \second{.930} & .938 & \second{.934} \\
  KEN & \second{.935} & .931 & .939 & \second{.935} & \second{.930} & .938 & \second{.934} & \second{.948} & .930 & .935 & .932 & .935 & .940 & .937 & .885 & .730 & .660 & .693 & .905 & .945 & .925 \\
  GSFND & .915 & .915 & .921 & .918 & .910 & .914 & .912 & .899 & .880 & .888 & .884 & .870 & .880 & .875 & .892 & .750 & .780 & .765 & .922 & .940 & .931 \\
  MViR & .913 & .918 & .904 & .911 & .907 & .919 & .913 & .917 & .926 & .902 & .914 & .912 & .923 & .917 & .884 & .753 & .629 & .685 & .913 & .951 & .932 \\
  \textbf{Ours} & \best{.939} & \best{.953} & .938 & \best{.945} & .929 & \best{.944} & \best{.936} & \best{.952} & \best{.965} & \best{.950} & \best{.957} & \second{.939} & \second{.954} & \best{.946} & \best{.908} & \best{.800} & \best{.810} & \best{.805} & \best{.935} & .943 & \best{.939} \\
  \bottomrule
  \end{tabular}
  \end{adjustbox}
  \endgroup
\end{table}

Table~\ref{tab:overall} reports overall results on Weibo, Weibo-21, and Gossip using the baselines from Section~\ref{subsec:baselines}. Three findings stand out. First, our route-then-decompose design achieves the best accuracy on all three benchmarks---for instance, 0.952 accuracy with 0.952 macro-F1 on Weibo-21 and 0.908 accuracy with 0.872 macro-F1 on Gossip---outperforming strong baselines such as InSide and KEN by 0.4--1.2\% in accuracy, and on the harder Gossip set it keeps Real F1 high (0.939) while lifting Fake F1 to 0.805, so coupling adaptive routing with explicit interaction decomposition helps most on the difficult class. Second, performance tracks how finely a method treats cross-modal interaction: simple fusion (EANN, SpotFake) trails, ambiguity- and consistency-aware methods (CAFE, CMC, BMR) improve, and recent expert-routing designs (MIMoE-FND, InSide, KEN) push the upper bound; by decomposing fusion into \emph{uniqueness}, \emph{redundancy}, and \emph{synergy} rather than a single fused vector or a scalar weight, IMEF advances this frontier, confirming that resolving heterogeneous interactions, not merely adding capacity, is what matters. Third, the model stays robust as the dominant modality shifts, keeping precision and recall closely matched for both classes (e.g., Weibo-21 Fake 0.965/0.950, Real 0.939/0.954) and gaining a stable 1.3 points in accuracy from Weibo to Weibo-21, because MMEG routes each post instance-wise and can lean on whichever modality is informative---a property we probe directly in Sec.~\ref{sec:robust_shift}.

\subsection{Ablation Studies}
\label{sec:ablation}

To dissect IMEX-FND and quantify each component's contribution, we evaluate the variants below; all keep the encoders, data splits, and optimization of the main experiments and change only the target component. Table~\ref{tab:ablation} reports Macro-F1 (averaged over Fake/Real F1) on Weibo, Weibo-21, and Gossip, with drops w.r.t.\ the full model in \textcolor{red}{red}.
\begin{itemize}
\item \textbf{w/o IMEF}: the final head uses only the summed streams $\mathbf{t}{+}\mathbf{v}{+}\mathbf{m}$, with no interaction decomposition.
\item \textbf{w/o MMEG}: each modality enters IMEF as its pooled encoder feature, not routed experts.
\item \textbf{w/o MR\,\&\,$L_{\text{int}}$}: the interaction experts train without modality replacement or the interaction losses.
\item \textbf{w/o Aux}: only the final classifier is kept, removing the per-modality heads $p_T,p_I,p_F$.
\item \textbf{SF}: a simple-fusion baseline without MMEG and IMEF, classifying the concatenated pooled features.
\end{itemize}

\begin{table}[t]
  \centering
  \caption{Ablation study of core components on three datasets, reporting Macro-F1. (Higher is better)}
  \label{tab:ablation}
  \small
  \renewcommand{\arraystretch}{1.15}
  \setlength{\tabcolsep}{4pt}
  \begin{tabular}{l|c|c|c}
    \hline
    \rowcolor{blue!10}
    \textbf{Variant} & \textbf{Weibo} & \textbf{Weibo-21} & \textbf{Gossip} \\
    \hline
    \rowcolor{yellow!12}
    \textbf{Full} & \textbf{0.940} & \textbf{0.952} & \textbf{0.872} \\
    \hline
    w/o IMEF & 0.920 (\textcolor{red}{-0.020}) & 0.930 (\textcolor{red}{-0.022}) & 0.849 (\textcolor{red}{-0.023}) \\
    \hline
    w/o MMEG & 0.928 (\textcolor{red}{-0.012}) & 0.942 (\textcolor{red}{-0.010}) & 0.858 (\textcolor{red}{-0.014}) \\
    \hline
    w/o MR\,\&\,$L_{\text{int}}$ & 0.932 (\textcolor{red}{-0.008}) & 0.945 (\textcolor{red}{-0.007}) & 0.861 (\textcolor{red}{-0.011}) \\
    \hline
    w/o Aux & 0.936 (\textcolor{red}{-0.004}) & 0.948 (\textcolor{red}{-0.004}) & 0.866 (\textcolor{red}{-0.006}) \\
    \hline
    SF & 0.908 (\textcolor{red}{-0.032}) & 0.921 (\textcolor{red}{-0.031}) & 0.832 (\textcolor{red}{-0.040}) \\
    \hline
  \end{tabular}
\end{table}

Removing IMEF causes the largest single-component drop on all datasets ($-0.020$/$-0.022$/$-0.023$ Macro-F1), confirming that explicitly decomposing heterogeneous interactions is the most important design; removing MMEG is next ($-0.012$/$-0.010$/$-0.014$), showing that instance-wise routing is what supplies robustness under dominant-modality shift. Disabling MR and $L_{\text{int}}$ also weakens Macro-F1, indicating that the perturbation signal drives experts to specialize rather than collapse into redundant copies, while the auxiliary heads add a small but stable gain. The ordering w/o IMEF $>$ w/o MMEG $>$ w/o MR\,\&\,$L_{\text{int}}$ $>$ w/o Aux mirrors the route-then-decompose logic: decomposition contributes most, adaptive routing second, each relying on its dedicated training signal.

\subsection{Robustness to Dominant-Modality Shift}
\label{sec:robust_shift}
To probe difficulty~(ii) directly, we test whether MMEG helps when the decisive modality varies across posts. Using the auxiliary text- and image-only heads ($p_T$, $p_I$), we split each test set into an agreement subset (the two unimodal predictions concur) and a conflict subset (they disagree); the conflict subset approximates posts whose dominant modality is ambiguous or shifts. Table~\ref{tab:shift} compares the full model with the $-$MMEG variant on both subsets.

\begin{table}[t]
  \centering
  \caption{Robustness under dominant-modality shift: Macro-F1 on the \emph{agreement} vs \emph{conflict} subsets (posts where the text- and image-only heads agree / disagree).}
  \label{tab:shift}
  \small
  \setlength{\tabcolsep}{5pt}
  \begin{tabular}{l cc cc cc}
    \hline
    \rowcolor{gray!15}
     & \multicolumn{2}{c}{Weibo} & \multicolumn{2}{c}{Weibo-21} & \multicolumn{2}{c}{Gossip} \\
     Variant & Agree & Conflict & Agree & Conflict & Agree & Conflict \\
    \hline
    Full        & 0.962 & 0.862 & 0.969 & 0.895 & 0.911 & 0.761 \\
    $-$MMEG     & 0.956 & 0.828 & 0.964 & 0.867 & 0.903 & 0.730 \\
    \hline
    $\Delta$    & $-$0.006 & $-$0.034 & $-$0.005 & $-$0.028 & $-$0.008 & $-$0.031 \\
    \hline
  \end{tabular}
\end{table}

Removing MMEG hurts the conflict subset far more than the agreement subset ($\Delta$Macro-F1 $\approx -0.031$ for conflict vs.\ $-0.006$ for agreement on average), confirming that instance-wise routing is the component that absorbs dominant-modality shift, whereas on easy, agreement posts the routing margin is naturally small.

\subsection{Traceability and Interaction Analysis}
\label{sec:traceability}
The interaction weights $\boldsymbol{\alpha}$ expose which interaction drives each decision, addressing the transparency side of difficulty~(i), and can be read at two levels. At the dataset level, aggregating $\boldsymbol{\alpha}$ over a test set yields an interaction profile: Gossip is dominated by text- and image-\emph{uniqueness} ($\bar{\alpha}=0.291$ and $0.268$), whereas Weibo leans most on \emph{synergy} ($\bar{\alpha}=0.308$), with Weibo-21 in between, matching their differing modality reliance. At the single-post level, $\boldsymbol{\alpha}$ attributes the verdict to one pattern---a claim with a mismatched photo to \emph{synergy}, a spliced image to image-\emph{uniqueness}, and a corroborating caption--image pair to \emph{redundancy}---as detailed in Table~\ref{tab:cases}, turning an opaque fusion score into a per-sample rationale.

\begin{table}[t]
  \centering
  \caption{Local interaction attribution for representative posts: per-sample IMEF weights $\boldsymbol{\alpha}$ over the five experts (dominant one in \textbf{bold}). The weight pattern matches the underlying text--image relation, and every prediction agrees with the ground-truth label.}
  \label{tab:cases}
  \footnotesize
  \setlength{\tabcolsep}{4pt}
  \renewcommand{\arraystretch}{1.15}
  \begin{adjustbox}{width=\linewidth}
  \begin{tabular}{llcccccc}
    \hline
    \rowcolor{gray!15}
    Dataset & Text--image relation & $\alpha_T$ & $\alpha_I$ & $\alpha_F$ & $\alpha_{\mathrm{syn}}$ & $\alpha_{\mathrm{red}}$ & Pred. \\
    \hline
    Weibo & Claim--photo mismatch & 0.08 & 0.07 & 0.10 & \textbf{0.62} & 0.13 & Fake\,\checkmark \\
    Gossip & Spliced/tampered image & 0.06 & \textbf{0.71} & 0.08 & 0.09 & 0.06 & Fake\,\checkmark \\
    Gossip & Exaggerated text, neutral image & \textbf{0.68} & 0.07 & 0.09 & 0.10 & 0.06 & Fake\,\checkmark \\
    Weibo-21 & Caption--image corroborate & 0.09 & 0.08 & 0.12 & 0.14 & \textbf{0.57} & Real\,\checkmark \\
    \hline
  \end{tabular}
  \end{adjustbox}
\end{table}

\subsection{Hyperparameter Settings}
\label{sec:hparams}

We study the sensitivity of the fusion weight $\lambda$, which controls the hybrid fusion between the conventional fused representation and the IMEF interaction feature. Specifically, $\lambda{=}0$ reduces the model to a conventional fusion path, while larger $\lambda$ places more emphasis on interaction-aware fusion. We sweep $\lambda \in \{0.0, 0.1, 0.3, 0.6, 1.0\}$ and report Accuracy together with class-wise F1 scores (Fake/Real).

\begin{figure}[h]
  \centering
  \includegraphics[width=\linewidth, page=1, trim=0 0 0 0, clip]{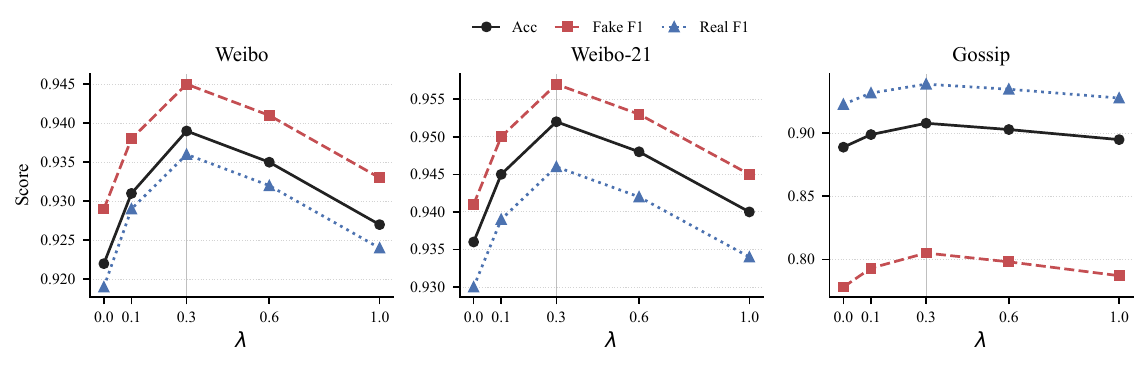}
  \caption{Sensitivity of $\lambda$ on Weibo, Weibo-21, and Gossip (left to right). Each subplot shows Accuracy and class-wise F1 scores (Fake/Real).}
  \label{fig:lambda_sweep}
\end{figure}

As Figure~\ref{fig:lambda_sweep} shows, a moderate interaction weight generally performs best across datasets. Performance improves from $\lambda{=}0$ to $\lambda{=}0.3$ and then slightly saturates or decreases, indicating that interaction-aware fusion helps but should be balanced with the conventional path: routing (MMEG) and decomposition (IMEF) are complementary, not substitutes. Unless otherwise specified, we use $\lambda{=}0.3$ in all experiments.

\subsection{Efficiency Analysis}
\label{subsec:efficiency}
We compare efficiency against architecturally comparable detectors---neural fusion and expert-based models trained on frozen encoders---under identical backbones and input sizes (197 tokens, $224{\times}224$); we omit the LLM- and knowledge-augmented methods (GSFND, KEN, InSide), whose cost is dominated by external generation or retrieval and is thus not comparable on trainable parameters. We report trainable parameters (excluding frozen encoders) and runtime on a single NVIDIA RTX 3090 with batch size 64 (AMP). As shown in Table~\ref{tab:efficiency}, IMEX-FND sits in the middle of the parameter range yet attains the best accuracy (Table~\ref{tab:overall}), giving a favorable cost--performance trade-off; in particular, its interaction regularization adds only limited training overhead since the extra perturbed passes are confined to the lightweight IMEF module and disabled at inference, keeping its test-time latency close to the most efficient baselines.

\begin{table}[t]
\centering
\caption{Efficiency comparison on Weibo-21 (batch size 64) against architecturally comparable baselines. Params counts trainable parameters only.}
\small
\setlength{\tabcolsep}{4pt}
\begin{tabular}{lccc}
\hline
\rowcolor{gray!15}
Method & Params (M) & Train (ms/iter) & Test (ms/iter) \\
\hline
SpotFake & 20.4 & 88 & 32 \\
EANN & 22.6 & 94 & 34 \\
CAFE & 25.8 & 103 & 37 \\
CMC & 27.3 & 108 & 38 \\
DAMMFND & 29.2 & 115 & 42 \\
FND-CLIP & 31.6 & 117 & 40 \\
\rowcolor{blue!8}
IMEX-FND (ours) & 33.5 & 121 & 43 \\
BMR & 36.8 & 127 & 45 \\
MIMoE-FND & 43.0 & 133 & 47 \\
\hline
\end{tabular}
\label{tab:efficiency}
\end{table}

\section{Conclusion}
We presented IMEX-FND, an interaction-aware framework that addresses the limited robustness and interpretability of multimodal fake news detection via a route-then-decompose design. A Multi-Modal Expert Gateway (MMEG) routes each modality through specialized and shared experts and builds a CLIP-grounded fusion stream, while an Interaction-aware Multi-Modal Expert Fusion (IMEF) disentangles cross-modal dynamics into \emph{uniqueness}, \emph{redundancy}, and \emph{synergy} via traceable sample-wise weights, turning an opaque fusion score into an inspectable account of modality contributions and interactions. Experiments on Weibo, Weibo-21, and Gossip show that IMEX-FND achieves state-of-the-art performance, surpassing competitive baselines by $0.4\% \text{--} 1.2\%$ while offering superior transparency with fewer parameters. Future work will explore causal reasoning~\cite{miao2026r3rec}. Efficient refinement under limited feedback is another direction~\cite{miao2026universalrefinementinteractionorderoptimal}.

\begin{credits}
\subsubsection{\ackname}
This work was supported by the National College Student Innovation and Entrepreneurship Training Program under Project No.~202619145026.
\end{credits}

\bibliographystyle{splncs04}
\bibliography{references}

\end{document}